\documentclass[twocolumn,a4paper]{article}
\usepackage[utf8]{inputenc}
\usepackage{fourier}
\usepackage{nimbussans}
\usepackage[cal=boondoxo]{mathalfa}
\usepackage{mathtools,graphicx,microtype,xcolor}
\usepackage[numbers,sort&compress]{natbib}
\makeatletter \def\NAT@def@citea{\def\@citea{\NAT@separator\,}} \makeatother
\definecolor{boxcolor}{cmyk}{1, 0.5, 0, 0.1}
\definecolor{linkscolor}{cmyk}{0.6, 0.3, 0, 0.9}

\usepackage[colorlinks=true,allcolors=linkscolor!60]{hyperref}

\usepackage{titlesec}
\usepackage{titling}
\titleformat*{\section}{\large \bfseries \sffamily}
\titleformat*{\subsection}{\large \bfseries \sffamily}
\titleformat*{\subsubsection}{\large \bfseries \sffamily}

\usepackage[centering,margin=1.5cm]{geometry}

\usepackage{tcolorbox}
\newtcolorbox{abstractbox}{
  arc=0pt,
  boxrule=0pt,
  colback=boxcolor!20,
  boxsep=1em,
  left=0pt, right=0pt, bottom=0pt, top=0pt,
  width=1\columnwidth
}

\renewenvironment{abstract}{
   \noindent
   \begin{minipage}{1\columnwidth}
   \upshape\sffamily 
   \begin{abstractbox}
   \fontsize{9}{14}\selectfont
  }{
   \end{abstractbox}
   \end{minipage} 
   \vskip 2.0em
  }

\makeatletter
 \def\@textbottom{\vskip \z@ \@plus 1pt}
 \let\@texttop\relax
\makeatother

\title{\textbf{Coulomb pairing of electrons in thin films with strong spin-orbit interaction}}

\author{Yasha Gindikin\thanks{Corresponding author: \href{mailto:gindikin@protonmail.ch}{gindikin@protonmail.ch}}\; and Vladimir A.\ Sablikov\\
\textit{\footnotesize Kotel'nikov Institute of Radio Engineering and Electronics,
Russian Academy of Sciences, Fryazino, 141190, Russia}}
\date{}

\begin{document}
\maketitle

\begin{abstract}
In low-dimensional structures with strong Rashba spin-orbit interaction (SOI), the Coulomb fields between moving electrons produce a SOI component of the pair interaction that competes with the potential Coulomb repulsion. If the Rashba SOI constant of the material is sufficiently high, the total electron-electron interaction becomes attractive, which leads to the formation of the two-electron bound states. We show that because of the dielectric screening in a thin film the binding energy is significantly higher as compared to the case of the bulk screening.
\end{abstract}

\section{Introduction}

Electronic mechanisms of electron pairing with high binding energy is a challenging problem that opens up broad prospects for the discovery of novel many-particle effects in various low-dimensional structures and modern materials, not to mention the high-temperature superconductivity~\cite{combescot2015excitons,kagan2013modern}. Recently we have proposed a purely electronic mechanism that potentially could provide a high enough binding energy~\cite{PhysRevB.98.115137,2018arXiv180410826G}. It is caused by a spin-dependent component of the electron-electron (e-e) interaction that appears because of the Rashba-like spin-orbit interaction (SOI) induced by the Coulomb field between electrons~\cite{PhysRevB.95.045138}. The origin of the spin-orbit component of the pair interaction of electrons is similar to that of the spin-dependent component of the impurity potential that causes skew scattering and side-jumping in the theory of the extrinsic spin Hall effect~\cite{Vignale2009}. This mechanism can be effective in materials with a strong Rashba SOI\@. The conditions under which electron pairs are formed, the bound-state spectrum and electronic structure were studied for the quantum wires and two-dimensional (2D) electron systems. For realistic conditions the binding energy was estimated to be in the meV range. In the present paper we show that the binding energy can be strongly increased by a suitable choice of the dielectric environment.
  
The pairing mechanism has unusual properties due to the key role that the SOI plays in the formation of the pairs. The SOI component of the e-e interaction depends on both the spin and momentum of electrons. Therefore the e-e interaction becomes attractive for a certain electron spin orientation tied to the momentum. This leads to the formation of the pairs of two distinct kinds with different spin structure depending on what type of motion creates the SOI: the relative motion of electrons or the motion of their center of mass. The binding energy of the electron pair is set by the SOI constant of the material, the magnitude of the electric field and its coordinate dependence.

In experiments, the 2D electron system is implemented in a thin film, the surrounding environment of which is known to strongly affect the electric field in the film. In a recent paper, we considered a 2D electron system embedded in a dielectric medium with the same dielectric constant as that of the material of the 2D layer with SOI\@. In this case the problem is solved analytically~\cite{PhysRevB.98.115137}, which allows us to prove the existence of the two-electron bound states, find their general properties and estimate the binding energy to be on the level of meVs.

However, from the point of view of the experimental implementation, of greater interest is the situation where the dielectric constant $\epsilon$ of the surroundings is much lower than that of the material with strong SOI\@. This situation is also interesting theoretically, since the presence of the low-$\epsilon$ surroundings leads not only to an increase in the interaction potential, but also to the significant change in its spatial dependence, especially at a small distance between the particles~\cite{rytova,keldysh1979coulomb}. The latter is especially important in our case, since the attractive component of the interaction caused by the SOI is determined by both the magnitude of the electric field and its coordinate dependence.

In this paper, we study the bound states in a thin film with strong SOI in a low-$\epsilon$ dielectric environment taking fully into account the dielectric screening. Such electronic systems are realized on the basis of graphene, 2D transition metal dichalcogenides, and thin layers of Bi\textsubscript{2}Se\textsubscript{3}. Although in such materials the band spectrum can be quite complex, in the present work we confine ourselves to a single-band model, which is nevertheless sufficient to capture the new effect of SOI\@. We find that the dielectric screening in the layer strongly facilitates the pairing to increase the binding energy by an order of magnitude.

\section{The model}
We start with a Hamiltonian of two interacting electrons in a layer situated in the $x$-$y$ plane. The kinetic energy is $H_{\mathrm{kin}} = (\mathbf{p}_1^2 + \mathbf{p}_2^2) /2m$,
where $\mathbf{p}_i= -i \hbar\nabla_{\mathbf{r}_i}$ is the momentum operator, $\mathbf{r}_i = (x_i,y_i)$ is the position of the $i$th electron, with $m$ being the effective electron mass. The layer width $d$ is assumed small, so that only one transverse subband is populated.

The \textit{e-e} interaction potential for a thin layer in vacuum is given by~\cite{rytova,keldysh1979coulomb}
\begin{equation}
\label{RK}
	U(\mathbf{r}) = \frac{\pi e^2}{2 r_0}\left[ H_0\left(\frac{r}{r_0}\right) - Y_0\left(\frac{r}{r_0}\right) \right]\,,
\end{equation}
with $H_0$ being the Struve function, and $Y_0$ being the Bessel function of the second kind~\cite{olver}. The screening length $r_0$ sets the crossover scale between the long-range $\sim 1/r$ Coulomb tail of the potential and its short-range logarithmic $\sim \log r$ divergence. The screening length can be estimated as $r_0 = \epsilon d/2$, with $\epsilon$ being the in-plane component of the dielectric tensor of the bulk material~\cite{PhysRevB.88.045318}.

The two-body SOI is given by~\cite{PhysRevB.98.115137}
\begin{equation}
\label{SOI}
	H_{\mathrm{SOI}} = \frac{\alpha}{\hbar} \sum_{i \ne j}  \left[E_y(\mathbf{r}_i - \mathbf{r}_j) p_{i x} - E_x(\mathbf{r}_i - \mathbf{r}_j) p_{i y} \right] \sigma_{z_i}\,,
\end{equation}
with $\sigma_{z_i}$ being the Pauli matrix, and $\alpha$ being the material-dependent SOI constant, which we assume positive for definiteness. The electric field, acting on the $i$th electron from the $j$th electron, is related to the Rytova-Keldysh potential of Eq.~\eqref{RK} via $\mathbf{E}(\mathbf{r}) = \frac{1}{e}\nabla U(\mathbf{r})$. Equation~\eqref{SOI} describes a two-particle interaction, which is attractive for a certain spin orientation locked to momentum.

The Schr\"odinger equation for the two-electron wave function $\Psi(\mathbf{r}_1,\mathbf{r}_2) = {\left(\Psi_{\uparrow \uparrow},\Psi_{\uparrow \downarrow},\Psi_{\downarrow \uparrow},\Psi_{\downarrow \downarrow}\right)}^{\intercal}$ splits into four uncoupled equations for the spinor components. 

Switch from the positions of the individual electrons to the relative position $\mathbf{r} = \mathbf{r}_1 - \mathbf{r}_2$ and the center-of-mass position $\mathbf{R} = (\mathbf{r}_1 + \mathbf{r}_2)/2$. Also introduce the corresponding momentum operators, $\mathbf{p} =  -i \hbar\nabla_{\mathbf{r}}$ and $\mathbf{P} =  -i \hbar\nabla_{\mathbf{R}}$.

The equations for $\Psi_{\uparrow \uparrow}$ and $\Psi_{\uparrow \downarrow}$ read as
\begin{equation}
\label{rel}
	\left [- \frac{\hbar^2}{m} \nabla^2_{\mathbf{r}} - \frac{\hbar^2}{4m} \nabla^2_{\mathbf{R}} + U(\mathbf{r}) + \frac{2\alpha}{\hbar} \frac{E(\mathbf{r})}{r} {(\mathbf{r} \times \mathbf {p})}_z \right ] \Psi_{\uparrow \uparrow}
	= \varepsilon_{\uparrow \uparrow} \Psi_{\uparrow \uparrow}
\end{equation}
and
\begin{equation}
\label{conv}
	\left[- \frac{\hbar^2}{m} \nabla^2_{\mathbf{r}} - \frac{\hbar^2}{4m} \nabla^2_{\mathbf{R}} + U(\mathbf{r}) + \frac{\alpha}{\hbar} \frac{E(\mathbf{r})}{r} {(\mathbf{r} \times \mathbf {P})}_z \right] \Psi_{\uparrow \downarrow} 
	= \varepsilon_{\uparrow \downarrow} \Psi_{\uparrow \downarrow}\,.
\end{equation}
The equations for $\Psi_{\downarrow \downarrow}$ and $\Psi_{\downarrow \uparrow}$ are obtained by changing the sign before $\alpha$ in the above equations, respectively.

Analysis shows that Eqs.~\eqref{rel} and~\eqref{conv} have solutions describing the bound states of electrons of different nature quite similarly to Ref.~\cite{PhysRevB.98.115137}.
We call the solutions of Eq.~\eqref{rel} that belong to the discrete part of the spectrum the relative bound states, since the effective electron attraction caused by the SOI is determined only by the relative motion of electrons. The solutions of Eq.~\eqref{conv} are called the convective bound states, because it is the motion of the electron pair as a whole that creates the SOI\@. Taking into account that the full solution of the system should be antisymmetric with respect to the particle permutation, we conclude that in the 2D system the relative bound states are triplet pairs, whereas the electrons with opposite spins are coupled in the convective bound state, which does not possess a definite spin~\cite{PhysRevB.98.115137}.

\section{Results}

Because of the translational invariance the wave functions can be written in the form $\Psi_{\uparrow \uparrow}(\mathbf{r},\mathbf{R}) = \exp (i \mathbf{K} \cdot \mathbf{R}) \psi_{\uparrow \uparrow}(\mathbf{r})$ and $\Psi_{\uparrow \downarrow}(\mathbf{r},\mathbf{R}) = \exp (i \mathbf{K} \cdot \mathbf{R}) \psi_{\uparrow \downarrow}(\mathbf{r},\mathbf{K})$.

First we consider the convective states, where the center-of-mass wave vector $\mathbf{K}$ affects the wave-function of the relative motion $\psi_{\uparrow \downarrow}(\mathbf{r},\mathbf{K})$ via the binding potential that equals
\begin{equation}
\label{cnbnd}
	 V(r,\phi) = \alpha E(r) K \sin \phi \,,
\end{equation}
 with $\phi$ being the polar angle measured from the $\mathbf{K}$-direction. The short-range asymptotics of the potential is 
 \begin{equation}
 \label{as}
	 V(r,\phi) \sim - \frac{Z e^2}{r} \sin \phi \,,
 \end{equation}
with the dimensionless SOI magnitude $Z = \alpha K/(e r_0)$. For sufficiently large $Z$, the binding potential of Eq.~\eqref{as} prevails over the weakly diverging repulsive potential $U(r) \sim \log (r/r_0)$ to allow for the bound states in the spectrum. 

It is interesting that owing to the dielectric screening in the layer, the attractive potential has a Coulomb-like form at small distance in contrast to the case of the bulk screening where the attractive potential diverges as $r^{-2}$. Therefore no regularization is needed to solve Eq.~\eqref{conv}.

Let us exploit a similarity to the Coulomb potential to make a crude estimate of the binding energy as $|\varepsilon| \propto Z^2 \cdot Ry$, the Rydberg constant in the material being $Ry = \hbar^2/2 m a_B^2$, with the Bohr radius $a_B = \epsilon \hbar^2/me^2$. Thus, the binding energy varies with the center-of-mass momentum as $|\varepsilon|\propto K^2$. We expect the size of the electron pair to be $\propto a_B/Z$.

Of course, the angular dependence of the binding potential makes a correction to this estimate. To account for this, we resort to numerical calculations with full potential of Eqs.~\eqref{RK} and~\eqref{cnbnd}. To be specific, assume $a_B = 100$~\AA, the layer thickness $d=0.2 a_B$, $\epsilon = 20$, and the dimensionless SOI constant $\tilde{\alpha} = \alpha/e a_B^2 = 1$, which is close to the parameters of such materials as Bi\textsubscript{2}Se\textsubscript{3}~\cite{manchon2015new}.

\begin{figure}[htb]
	\includegraphics[width=0.9\linewidth]{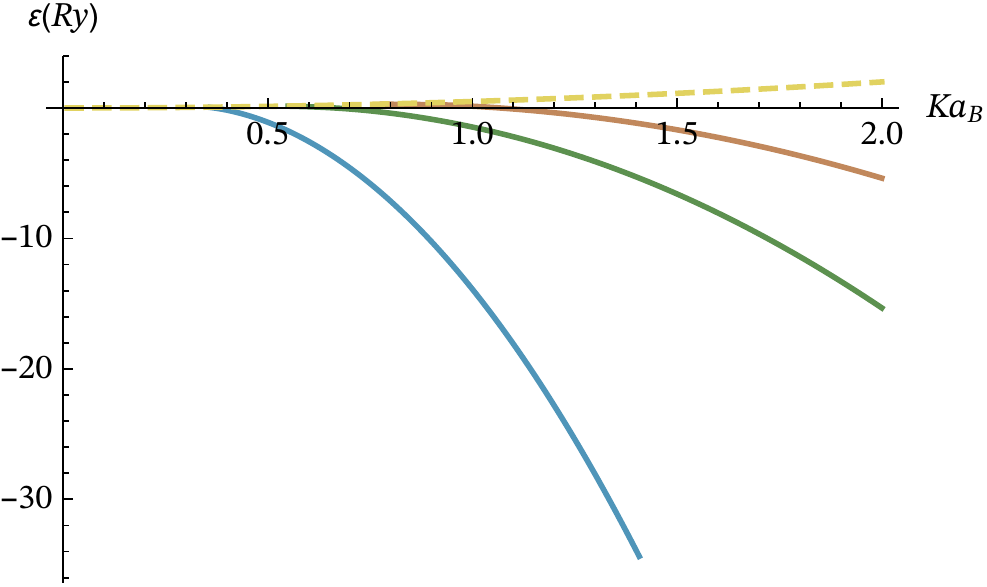}
		\caption{The system energy levels (solid lines) and the kinetic energy of the center of mass (dashed line) vs $K a_B$.}
	\label{fig1}
\end{figure}

Figure~\ref{fig1} shows the energies of the three lowest-lying convective states, with the kinetic energy of the center of mass included, as a function of the center-of-mass momentum. In other words, this is the energy dispersion of the convective electron pair. At the respective critical value of $K$, each bound state appears in the spectrum, with the binding energy growing approximately like $K^2$, in accordance with the above estimate.

Taking into account dielectric screening in the layer, the binding energy increases by a factor of about $\epsilon$ compared to that found in Ref.~\cite{PhysRevB.98.115137}, i.e.\ by an order of magnitude. Also note the SOI-induced renormalization of the effective mass of the electron pair, which even becomes negative. 

\begin{figure}[htb]
	\includegraphics[width=0.9\linewidth]{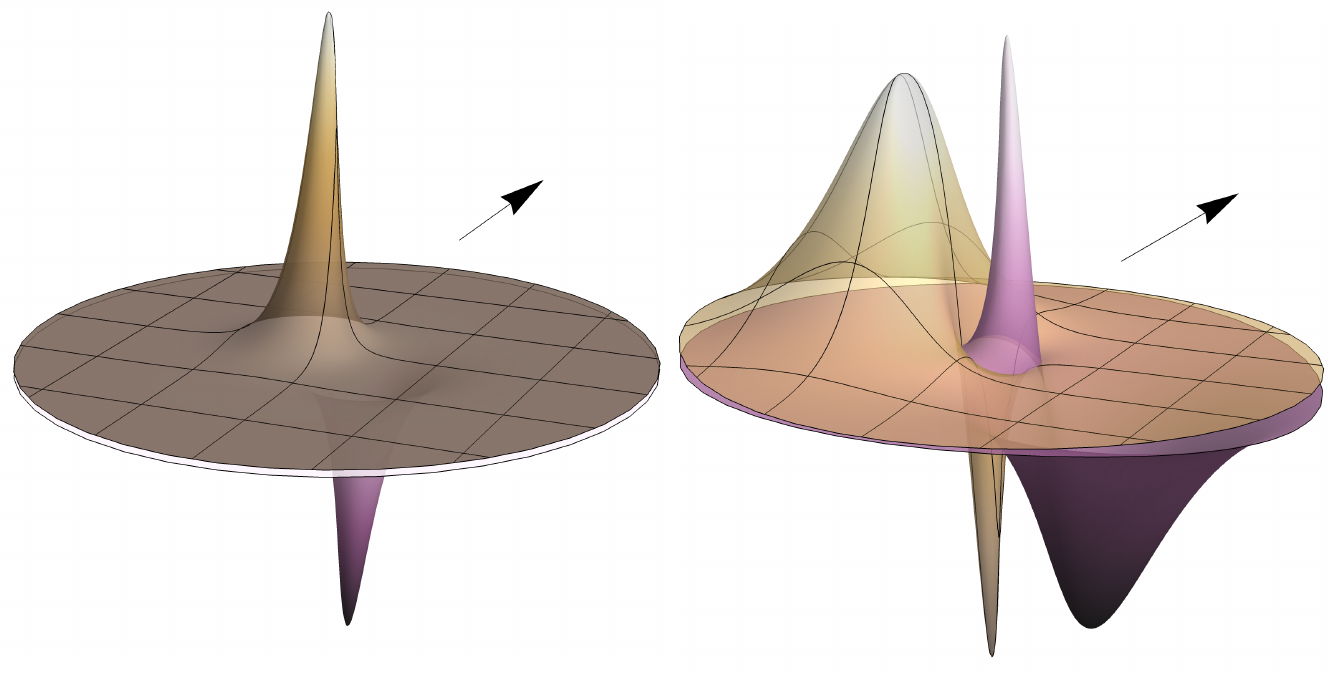}
		\caption{The spinor components of the convective state wave function for the ground state (left) and first excited state as functions of relative coordinates. The arrows show the direction of vector $\mathbf{K}$.}
	\label{fig2}
\end{figure}

Figure~\ref{fig2} shows the wave function of two lowest-lying convective states. Two surfaces, shown in different color in each figure, are the two spinor components $\psi_{\uparrow \downarrow}(\mathbf{r},\mathbf{K})$ and $\psi_{\downarrow \uparrow}(\mathbf{r},\mathbf{K})$. Note the strong angular dependence of the solutions, which is due to the highly anisotropic binding potential of Eq.~\eqref{cnbnd}.

Turning to the relative bound states, we note that since the orbital angular momentum along the $z$ direction $l_z = - i \partial_{\phi}$ commutes with the Hamiltonian, the wave function of the relative motion can be chosen as the eigenfunction of $l_z$, $\psi_{\uparrow \uparrow}(\mathbf{r}) = u(r) e^{i l \phi}$. The antisymmetric properties of $\Psi_{\uparrow \uparrow}$ require that the orbital angular quantum number $l$  be an odd integer.

The binding potential for the relative states is thus 
\begin{equation}
\label{relsing}
	V(r) = 2 \alpha l \frac{E(r)}{r}\,.
\end{equation}
Depending on the sign of $l$, this term can be repulsive or attractive. The relative bound state $\Psi_{\uparrow \uparrow}$ is supported by $l < 0$, and $\Psi_{\downarrow \downarrow}$ by $l > 0$. The spin projection of the relative state is seen to be locked to the orbital angular momentum. In what follows, we consider the case of $|l| = 1$ to minimize the centrifugal barrier $\propto l^2$.

The binding potential behaves as
\begin{equation}
\label{pat}
	V(r) \sim - \frac{2 \alpha}{r_0}\frac{e}{r^2}
\end{equation}
at small $r$. The \textit{e-e} attraction overcomes not only the much weaker $\propto \log (r/r_0)$ potential of repulsion, but also prevails over the centrifugal potential as long as $\tilde{\alpha} > 3 d/16 a_B$. This condition holds in our case.

The attractive $-1/r^{2}$ potential in Eq.~\eqref{pat} is a transitional singular potential that has been exciting interest for decades~\cite{RevModPhys.43.36}, not least because of its ubiquity in quantum physics. The inverse square potential appears in the three-body problem in nuclear physics~\cite{Efimov:1971zz}, it describes the point-dipole interactions in molecular physics~\cite{PhysRev.153.1} and the attraction of atoms to a charged wire~\cite{PhysRevLett.81.737}. Meanwhile, it has produced a lot of controversy when used with the Schr\"odinger equation. The requirement that its solutions are square integrable does not define a discrete orthogonal set of eigenfunctions with its eigenvalues; bound states with arbitrary energy $\varepsilon <0$ are possible. Imposing the orthogonality of the eigenfunctions does lead to a discrete spectrum of bound states that is nonetheless unbounded below, so there is no ground state~\cite{PhysRev.80.797}. This is interpreted as a fall to the center~\cite{landau1958course}. The problem is that the Hamiltonian is symmetric but not self-adjoint~\cite{Meetz1964}. To fix the problem, a number of regularization techniques was developed~\cite{PhysRevLett.85.1590,PhysRevA.64.042103,PhysRevA.76.032112}, which are essentially based on introducing a short-distance cut-off~\cite{PhysRevD.48.5940}.

The cut-off should be considered as a phenomenological parameter, the value of which can not be determined within the model considered, unless some outer mechanisms are taken into consideration or e.g.\ scaling-invariance requirements are imposed. A possible mechanism of cutting off the binding potential at small $r$ is related to the Zitterbewegung of electrons in crystalline solids~\cite{0953-8984-23-14-143201}, which leads to the cut-off $a$ that may actually be of the order of the film thickness $d$ or even larger. By cutting the potential of Eq.~\eqref{pat} at $r=a$ and imposing the zero boundary condition for the solution, we obtain the following estimate for the binding energy of the lowest-lying relative state,
\begin{equation}
	|\varepsilon| = \frac{2 Ry}{(a/a_B)^2} x_1^2\left(\sqrt{\frac{4 \tilde{\alpha}}{d/a_B}}\right)\,,
\end{equation}
where $x_1 (\mu)$ is the first (largest) zero of the Macdonald function $\mathcal{K}_{i \mu}(x)$~\cite{olver}. This gives the $|\varepsilon|$ magnitude of tens of Rydberg for the parameters considered.

\section{Conclusion}
We studied the Coulomb mechanism of electron pairing in low-dimensional structures with a strong Rashba SOI in the case where the e-e interaction is not screened by the environment. This situation is realized in recent experiments on freely suspended 2D structures~\cite{ROSSLER2010861,doi:10.1063/1.5019906}. It attracts growing interest because in this case the e-e interaction effects should be more pronounced. We have found that dielectric screening in the film crucially affects the pairing conditions and binding energy, which is increased by an order of magnitude as compared to the previously considered case of the bulk screening.

This work was partially supported by Russian Foundation for Basic Research (Grant No 17--02--00309) and Russian Academy of Sciences.

{\small \bibliography{paper}}

\end{document}